# People Are Highly Cooperative with Large Language Models, Especially When Communication Is Possible or Following Human Interaction

Paweł Niszczota[1,3,*], Tomasz Grzegorczyk[1], Alexander Pastukhov[2,3]

[1] Poznań University of Economics and Business, Humans & AI Laboratory (HAI Lab), Institute of International Business and Economics, Poznań, Poland

[2] University of Bamberg, Department of General Psychology and Methodology, Bamberg, Germany

[3] These authors contributed equally.

[*] Corresponding author: Paweł Niszczota (pawel.niszczota@ue.poznan.pl), Poznań University of Economics and Business, al. Niepodległości 10, 61-875 Poznań, Poland; https://orcid.org/0000-0002-4150-3646

**Acknowledgements**

P.N. acknowledges support from the National Science Centre, Poland (2021/42/E/HS4/00289).

We have benefited from discussions and feedback from Agne Kajackaite, Michał Białek, Rainer Rilke, and participants of the Centre for Behavioural and Experimental Social Science seminar.

**Author Contributions**

P.N., T.G. and A.P. conceived the study. P.N. and A.P. designed the experiments, collected data, and conducted data analyses. P.N., A.P. and T.G. wrote the manuscript. P.N. acquired the funding and supervised the work. P.N. and A.P. contributed equally to the work.

**Competing Interests**

The authors declare no competing interests.

**Data and Code Availability**

Data, code, and materials are available at https://osf.io/mq5ud/?view_only=bfda814d94a74d95a18a9057fd4fa36a.

Experiment 1 was pre-registered at https://aspredicted.org/ZH7_2K8 and Experiment 2 at https://aspredicted.org/SG5_FQF.

# People Are Highly Cooperative with Large Language Models, Especially When Communication Is Possible or Following Human Interaction

**Abstract**

Machines driven by large language models (LLMs) have the potential to augment humans across various tasks, a development with profound implications for business settings where effective communication, collaboration, and stakeholder trust are paramount. To explore how interacting with an LLM instead of a human might shift cooperative behavior in such settings, we used the Prisoner's Dilemma game—a surrogate of several real-world managerial and economic scenarios. In Experiment 1 ($N$=100), participants engaged in a thirty-round repeated game against a human, a classic bot, and an LLM (GPT, in real-time). In Experiment 2 ($N$=192), participants played a one-shot game against a human or an LLM, with half of them allowed to communicate with their opponent, enabling LLMs to leverage a key advantage over older-generation machines. Cooperation rates with LLMs—while lower by approximately 10–15 percentage points compared to interactions with human opponents—were nonetheless high. This finding was particularly notable in Experiment 2, where the psychological cost of selfish behavior was reduced. Although allowing communication about cooperation did not close the human-machine behavioral gap, it increased the likelihood of cooperation with both humans and LLMs equally (by 88%), which is particularly surprising for LLMs given their non-human nature and the assumption that people might be less receptive to cooperating with machines compared to human counterparts. Additionally, cooperation with LLMs was higher following prior interaction with humans, suggesting a spillover effect in cooperative behavior. Our findings validate the (careful) use of LLMs by businesses in settings that have a cooperative component.

Key words: large language models, GPT, cooperation, human-AI collaboration, Prisoner's Dilemma Game

# 1. Introduction

Large language models (LLMs) have become ubiquitous in the public space since the end of 2022, when ChatGPT was introduced (Vaswani et al. 2017, OpenAI 2022). Machines driven by LLMs have the potential to replace more and more interactions between humans: instead of doing some cognitively or psychologically taxing task by themselves, humans may increasingly decide that they would prefer an LLM to do this task on their behalf. However, to properly assess the effects of the rapid spread of LLMs in society, it is necessary to assess how human behavior is different when interacting with an LLM instead of a person. We use a prominent social dilemma to investigate how people change their behavior when their partner is an LLM and not another human. This serves as a crucial complement to the existing research, that uncovered the behavioral patterns of LLMs (Chen et al. 2023, Mei et al. 2024) but very rarely extensively compares human-LLM and human-human interactions (Rahwan et al. 2019, Dvorak et al. 2025).

In a social dilemma, one solution is better for people as a whole, while the other maximizes the short-term benefits of the individual, but at the expense of others. Our focus is on the Prisoner's Dilemma Game, in which cooperation leads to larger rewards for both players, yet defecting is a dominant strategy, as it guarantees an outcome that is better or at least equal to that of the opponent. Fortunately, cooperation is nonetheless common (Fehr and Schurtenberger 2018), as people recognize the benefits of cooperation, the costliness of non-cooperation, or both. However, the urge to behave pro-socially might be altered when interacting with a machine. There are at least two aspects of the interaction that might be different for human versus machine opponents. The first one is relevant to machines as a whole. The second one is only relevant to LLMs.

First, it is argued that cooperation is a social norm (Fehr and Schurtenberger 2018): people know that there is an expectation that one should cooperate, and that deviations from this norm are psychologically costly. More specifically, there appears to be a norm of conditional cooperation: cooperate if you expect that the *other person* will also cooperate. However, prior research suggests that the psychological cost of not following this social norm might be lower if the opponent is a machine (Fehr and Schurtenberger 2018). When defecting while playing against a machine, people might anticipate less disapproval from *other* humans, and no disapproval from the machine itself, as machines are perceived to lack emotions. A machine cannot feel betrayed when a person defects and, equally, cooperation will not make a machine feel better about a person. Also, people understand that an accumulated reward, whether abstract points or actual money, has no value for the machine opponent unless they believe that it is acting on behalf of a human.

Second, cooperation increases when players can communicate before making their individual decisions (Sally 1995, Fehr and Schurtenberger 2018). Until recently, proper communication with a machine opponent was impossible. However, after the arrival of LLMs, the two constraints listed above may no longer apply. The human-like ability to converse with people means that machines could be perceived as more similar to human beings, making social norms more relevant. Machines' ability to communicate provides an opportunity to influence their human opponents and agree on a common cooperation strategy. Thus, the main question of this research is to assess the extent to which the human-like appearance of an LLM will result in them being treated more in a way that they would be treated if a human was their opponent.

Prior research by Ishowo-Oloko and colleagues (Ishowo-Oloko et al. 2019) showed that fewer participants cooperated when playing a Prisoner's Dilemma Game against a machine rather than a human (Ishowo-Oloko et al. 2019). We conducted two experimental studies to understand whether LLMs have closed this human-machine behavioral gap in cooperation. In contrast to this and other studies investigating how people play against bots and bots-driven by LLMs (Dvorak et al. 2025), we use a within-subjects design, which allowed us to assess how *each* participant behaves differently when having to decide whether to cooperate or not. Experiment 1 followed the multiple-round Prisoner's Dilemma game design of Ishowo-Oloko and colleagues. In addition to human- and bot- (a specialized program) opponents, we added a third opponent: an LLM (GPT). Due to the relative novelty of this technology at the time the study was conducted, we allowed participants to interact with the LLM via text chat before playing the actual game, enabling them to experience its human-like conversational abilities. We expected that when playing against an LLM, participants' willingness to cooperate would lie somewhere in between the specialized program and the human opponent. This could mean that humans perceive LLMs as more human-like than older-generation machines.

However, in a repeated Prisoner's Dilemma game, the decision in the first round conflates prosocial and strategic decision-making. Even if participants had no reservations about defection when playing against a machine, they still had to take into account the inevitable retaliation that would compromise their game very early on. We addressed this issue in Experiment 2, where participants played a one-round Prisoner's Dilemma game against each opponent, rendering any strategic thinking irrelevant and putting self-interest at the forefront. More importantly, we manipulated the ability to communicate with the opponent: half of the participants had the opportunity to communicate and half did not. This allowed us to test the null hypothesis that permitting communication has a similar effect when playing against a human and an LLM.

In Section 2 we provide a brief overview of the literature that (1) discusses how LLMs can potentially enhance productivity in businesses, but also (2) how people might be less cooperative towards LLMs than humans, undermining these benefits. In Sections 3 and 4, we present the results of Experiment 1 and Experiment 2, respectively. In Section 4 we discuss our findings, also pointing to some directions for future research. In Section 5 we conclude our study.

## 2. Related Literature

### 2.1. The Value of Human-LLM Collaboration in Business

The integration of LLMs into business operations is increasingly recognized as a potential source of competitive advantage (Fui-Hoon Nah et al. 2023, Burton et al. 2024, Hao et al. 2024, Przegalinska et al. 2025). LLMs are already making significant inroads across diverse business functions, including customer service, marketing, office automation, IT operations, HR management, recruitment, banking, education, market research, training, logistics, finance, and innovation management (Budhwar et al. 2023, Srivastava et al. 2024, Mariani and Dwivedi 2024, Kshetri et al. 2024, Chiarello et al. 2024, Hermann and Puntoni 2024). Numerous organizations, such as Mastercard, Instacart, and Zalando in customer service, and Coca-Cola, Duolingo, and Salesforce across broader operations like product innovation and community building, have begun implementing these technologies (Brown et al. 2024). This widespread adoption suggests a trajectory toward even broader economic shifts, with projections indicating LLMs will significantly impact the labor market by altering tasks across various occupations and wage levels (Eloundou et al. 2024).

Incorporating LLMs into human-centric workflows offers substantial benefits, enhancing strategic planning, decision-making, innovation, and creative problem-solving, which can lead to improved financial, strategic, and environmental outcomes (Holmström and Carroll 2024, Boussioux et al. 2024, Burton et al. 2024). A key approach is the "human-in-the-loop" design, where humans actively guide or intervene in AI-driven processes, fostering collaboration towards shared objectives (Sowa et al. 2021, Shneiderman and Shneiderman 2022). This synergy leverages the complementary strengths of humans and machines: humans contribute contextual understanding, ethical judgment, and nuanced creativity, while LLMs provide speed, scalability, and computational power (Bilgram and Laarmann 2023). Positive impacts have been observed across various task types, including routine, complex, and creative work (Eloundou et al. 2024), although the effectiveness of this collaboration is often moderated by the individual's experience and expertise (Zhang et al. 2025).

Significant productivity gains are a major driver of LLM adoption. For example, Brynjolfsson et al. (2025) found that AI assistance increased worker performance by an average of 15%, with less-experienced workers realizing gains up to 30%. This study also noted improvements in task efficiency, reduced employee attrition, and enhanced skill development. Supporting these findings, Bastola et al. (2024) reported workload reductions through an LLM-powered smart reply system. Furthermore, human-AI teams have been shown to outperform human-only teams in complex scenarios, such as crisis-related resource allocation (McNeese et al. 2021). The concept of a "jagged technological frontier," where AI excels in specific areas that complement human skills, was illustrated by Dell'Acqua et al. (2023), who observed significantly higher productivity among consultants using GPT-4.

Beyond sheer productivity, LLMs demonstrate considerable potential in improving decision-making quality (Przegalinska et al. 2025). Research suggests LLMs can reduce cognitive load and mitigate heuristic biases by offering data-driven insights and supporting more analytical System 2 thinking (Kahneman 2011, Hao et al. 2024). In group settings, LLMs have been used effectively as "devil's advocates" to enhance decision accuracy (Chiang et al. 2024). These decision-support capabilities are finding application across various industries. Examples include assisting with financial analytics and investment strategies (Fatouros et al. 2024), informing behavioral economics experiments (Horton 2023), analyzing economic games (Aher et al. 2023), and improving supply chain risk management and contract intelligence (Srivastava et al. 2024). AI has also shown promise in overseeing the design processes of human engineering teams (Gyory et al. 2021).

Marketing and customer service represent another area ripe for human-LLM hybridization. LLMs are being integrated into chatbots, content creation, and predictive analytics to enhance customer experiences and marketing effectiveness (Brown et al. 2024, Saurith Moreno et al. 2024). These tools help improve personalization, communication efficiency, and consumer engagement. Studies indicate that hybrid human-LLM teams can generate more insightful marketing research outcomes compared to either humans or LLMs working alone (Arora et al. 2025). While LLMs perform well on routine marketing tasks like persona generation, necessitating workforce reskilling, their utility in highly creative tasks like product naming may be more nuanced, sometimes showing no clear advantage based on AI proficiency alone (Arora et al. 2025). However, for non-experts, LLM feedback can improve outputs like advertising copy (Chen and Chan 2024). AI-generated marketing communications have even achieved significantly higher click-through rates in some contexts (Angelopoulos et al. 2024). In customer relations, LLM adoption has helped consumers achieve better outcomes when lodging financial complaints, promoting fairness (Shin et al. 2025).

In summary, human-LLM collaboration offers substantial strategic, operational, and customer-facing advantages. By strategically implementing LLMs, organizations can harness hybrid intelligence to boost productivity, refine decision-making, and enrich consumer interactions. These collective findings underscore the significant value of human-LLM cooperation while highlighting the need for continued multidisciplinary research into the complexities of effective human-machine partnerships.

## 2.2. Challenges Hindering Human-LLM Cooperation

Despite the numerous benefits, realizing the full potential of human-LLM collaboration depends critically on the ability of both humans and AI agents to work together effectively (Kolbjørnsrud 2024). For humans, the willingness to cooperate with AI is heavily influenced by trust, which itself is linked to perceptions of transparency, consistency, and anthropomorphism (Glikson and Woolley 2020). However, significant barriers exist. The inherent lack of transparency in many complex AI systems (Ishowo-Oloko et al. 2019) and the documented potential for LLMs to generate inaccurate or misleading information (Fui-Hoon Nah et al. 2023) contribute to a tendency for people to view AI as less trustworthy and cooperative compared to fellow humans (Dvorak et al. 2025).

### 2.2.1. The Transparency Dilemma

One debated approach to fostering interaction has been to obscure the AI's identity, making users believe they are interacting with a human, as seen in applications like Google Duplex (O'Leary 2019). However, such deception raises ethical concerns and is unlikely to be a sustainable long-term strategy. This brings the challenge of fostering genuine human willingness to cooperate with identifiable AI systems to the forefront. A major obstacle is the "black box" nature of many complex algorithms; decisions made without clear rationale can erode trust, mirroring how unexplained actions damage interpersonal relationships (Burt 2019, Wang et al. 2024). Explainable AI (XAI) offers a potential solution, particularly through intention-based explanations that clarify the reasoning behind AI outputs, which can bolster trust and facilitate cooperation (Ai et al. 2021, Berberian et al. 2023).

### 2.2.2. Accountability and Anthropomorphism

Human cooperation is often grounded in social norms like reciprocity and fairness, which typically regulate interactions between people (Fehr and Schurtenberger 2018). Applying these norms to AI interactions may depend on anthropomorphism – the extent to which humans attribute human-like qualities to AI. If AI is not perceived as sufficiently "human-like," these social norms may seem less relevant, potentially increasing the likelihood of opportunistic or unfair behavior (Cohn et al. 2022). Compounding this, research indicates that people tend to hold AI agents less accountable for unfair actions compared to human counterparts, which can reduce the motivation to enforce fairness norms when interacting with AI (Zhang et al. 2025). Consequently, endowing LLMs with certain human-like traits – such as conversational fluency, emotional expression (Chandra et al. 2022), or even verbal and non-verbal cues like humor or empathy – might enhance their persuasiveness and encourage more cooperative interactions by making social norms feel more applicable (Jiang et al. 2024).

## 3. Experiment 1: How Differently Do People Behave Differently When Playing a Repeated Prisoner's Dilemma Game Against an LLM?

To perform an assessment of the cooperative behavior towards LLMs – relative to both humans and conventional bots (Ishowo-Oloko et al. 2019) – we conducted a within-subjects experiment, in which people played the Repeated Prisoner's Dilemma Game across various opponents.

### 3.1. Procedure

Participants played a thirty-round Repeated Prisoner's Dilemma game against 1) a *human* opponent, 2) a *bot* (a computer program developed specifically to play the Prisoner's Dilemma game), and 3) an *LLM*. When playing against a human opponent, the participants were split into pairs based on their arrival time. When playing against the LLM, participants were first given a chance to interact with it and they had to send 10 messages in total and received 10 answers.

In each round, participants had to decide whether to cooperate with the opponent or defect. Following their own decision, they were informed about the opponent's decision, payoffs, and their total accumulated points.

We adopted a similar incentive structure to Ishowo-Oloko et al. (2019). The flat fee of £2.50, which was based on the expected duration of the study, and intentionally corresponded to the minimum wage on Prolific, to engage participants as much as possible in the games. In addition, participants were informed that they would obtain a bonus payment based on the results of the economic game, up to £13.50.

Two main models were implemented to characterize participants' responses over the entire Repeated Prisoner's Dilemma Game, assuming parameters could co-vary depending on the opponent type. The first model was a Binomial Generalized Linear Model (GLM). The second model was an Ideal-Observer Model. The latter approach modeled decisions as arising from a weighted average of dynamically adjusted expectations about the player's own willingness to cooperate and the opponent's willingness to cooperate. More details can be found in the *Models of Cooperative Behavior* subsection of Appendix A.

The study was pre-registered at https://aspredicted.org/ZH7_2K8. Data, code, and materials are available at https://osf.io/mq5ud/?view_only=bfda814d94a74d95a18a9057fd4fa36a.

### 3.2. Results

The proportion of participants who cooperated in each round is shown in **Figure 1A** (see also **Figure S1** in Appendix B), along with posterior predictions of the ideal observer model. Over the course of the entire game, participants cooperated the most when playing against *human* participants (the proportion of rounds when players cooperated was $P_{cooperation}$ = 0.72 [0.70-0.74], mean and 97% exact binomial confidence interval), and the least when playing against the *LLM* ($P_{cooperation}$ = 0.34 [0.33-0.36]), with cooperation with the non-LLM *bot* being somewhere in-between ($P_{cooperation}$ = 0.59 [0.57-0.61]). However, it is important to differentiate between decisions made during the early and later rounds. Plausibly, behavior in early rounds is strongly influenced by expectations of the opponent's behavior, as the behavior itself has not been observed yet. Decisions made in later rounds, on the other hand, are influenced by the decisions and outcomes of prior rounds.

The dependence of decisions on prior experience is shown in **Figure 1B** (see also **Figure S2** in Appendix B) which depicts the total points accumulated by participants and their machine opponents over game rounds. It is clear that the *bot* opponent was too cooperative (or too lenient to opponents deviating from conditional cooperation; Fehr and Schurtenberger 2018) and exposed itself to exploitation, as participants outperformed it throughout the entire game. In contrast, the *LLM* played aggressively (quickly deviated from the conditional cooperation norm). This produced a short-term advantage, as participants adapted to the LLM's strategy around Round 15 on average (before this point they were consistently outplayed), and they managed to be on par with the LLM only by the end of the game. The consequence of this strategy was a lower overall point haul for both parties.

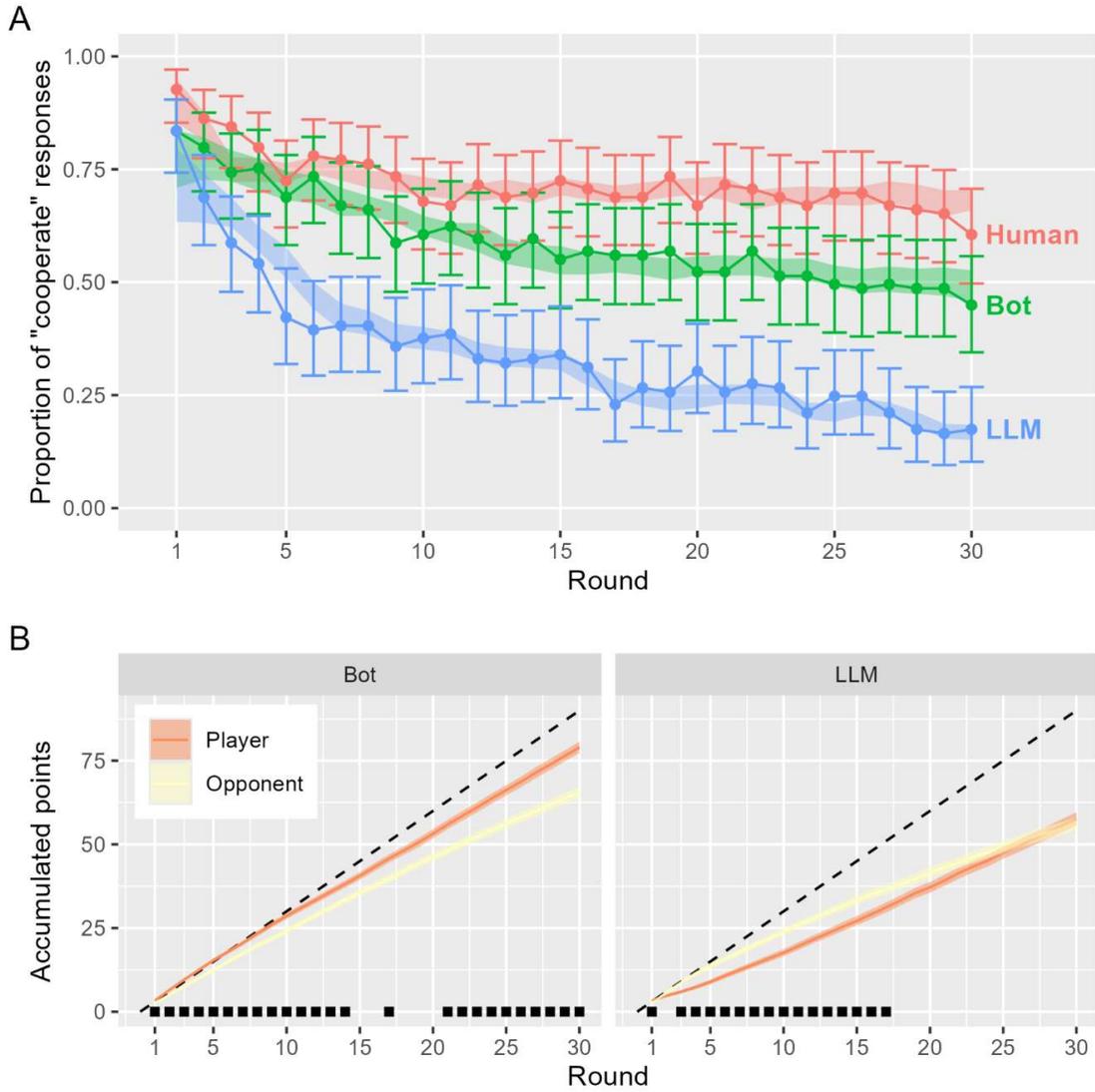

**Figure 1. Cooperation and Accumulated Points Over All Game Rounds in Experiment 1**
*Notes.* (A) The proportion of cooperative responses when playing against a specific opponent on each round for the binomial GLM model. Circles and error bars correspond to mean values and bootstrapped non-parametric percentile 97% confidence intervals, respectively. Stripes depict the 97% credible interval for posterior predictions. (B) Accumulated points for the human player and their machine opponent. Lines and stripes correspond to mean and non-parametric bootstrapped 97% confidence interval. The dashed line shows the accumulated points if both opponents would always cooperate.

In contrast to most game rounds, the first two rounds primarily reflect the participants' initial predisposition to cooperate with their opponents. We quantified the difference in cooperation rates using a binomial multi-level model with independent probabilities for each round, opponent type, and prior response and participant identity as pooled random factors. The results, presented in **Table 1**, show that participants were less cooperative with both the *bot* and the *LLM* compared to the *human* opponent from the very first round. As noted earlier, they did not differentiate between the two machine opponents in the first round.

**Table 1. Participants' Decision in the First Two Rounds of Experiment 1**

| Round | Decision on previous round | | The proportion of participants who cooperated | | |
|---|---|---|---|---|---|
| | Player | Opponent | Opponent: Human | Opponent: Bot | Opponent: LLM |
| 1 | – | – | 0.93 [0.86, 0.97] | 0.83 [0.75, 0.90] | 0.83 [0.75, 0.90] |
| 2 | C | C | 0.89 [0.81, 0.95] | 0.91 [0.83, 0.96] | 0.83 [0.74, 0.90] |
| | C | D | 0.67 [0.22, 0.96] | – | 0.50 [0.01, 0.99] |
| | D | C | 0.80 [0.28, 0.99] | 0.22 [0.06, 0.48] | 0.00 [0.00, 0.19] |
| | D | D | 0.33 [0.01, 0.91] | – | – |

*Notes.* This table shows the probability of cooperation in Rounds 1 and 2 given the player's and opponent's decisions in Round 1. Values represent means and binomial 97% confidence intervals. Cells with missing values mean that this combination never occurred in the data. See also **Table S1** in Appendix B for information on the statistical significance of the difference between pairs of opponents. C = *Cooperation*, D = *Defection*.

In Round 2, participants cooperated less with the *LLM* regardless of the previous choices made by both players. For example, even when both players cooperated in the first round, participants were less likely to cooperate again when playing against the *LLM* than against a fellow *human* or the *bot*. This difference is even more pronounced when participants defected, but the opponent cooperated in Round 1. In such cases, participants played very aggressively against both machine opponents, with less cooperation with *LLM*. This pattern could reflect a defense against expected retaliation, or an attempt to exploit the cooperative nature of the machine opponents.

To summarize, we found that initially (in the first round), participants cooperated more readily with human opponents than with the machine opponents, and that they did not differentiate between the *LLM* and the *bot*. In the following rounds, they cooperated significantly less with the *bot* and even less with the *LLM*. However, this might be largely due to the difference of strategies used by the two machine opponents.

## 4. Experiment 2: How Do People Play Against an LLM in a One-shot Prisoner's Dilemma Game, and Does Pre-game Communication Change This Behavior?

Experiment 1 showed that participants' predisposition during the first round was to cooperate less when playing against machine opponents. However, initially, they did not differentiate between both machine players: the *LLM* and a specialized *bot*. On the one hand, this is surprising, as it runs counter to the expectation that the human-like communication capabilities of LLMs make them appear more human-like, elevating the likelihood of cooperation with them relative to more conventional machines. On the other hand, the content of the communication could have rendered it less effective as it had no bearing on the game (Dawes et al. 1977). In other words, even if the player discussed a common strategy with the LLM, it would not be used as context for the LLM's decision. In addition, the initial decision in Experiment 1, as well as in prior work (Ishowo-Oloko et al. 2019), was a necessary compromise between prosocial and strategic decisions, making it hard to elucidate the former. We addressed these issues in Experiment 2, where participants played the classic one-round version of the Prisoner's Dilemma game against another *human* and an *LLM*.

### 4.1. Procedure

In Experiment 2, 192 participants played a one-shot Prisoner's Dilemma game. They were split into four groups (48 participants in a group) based on a 2 × 2 condition design. The groups differed in (1) whether they were able to communicate with an opponent before the actual game (*Communication* and *No communication* conditions) and (2) the order of opponents (human opponent followed by an LLM opponent labelled as *Human-LLM*, or LLM followed by a human opponent labelled as *LLM-Human*). Before each single-round game, participants were informed about their opponent (a human or an LLM). In the *Communication* condition, they were able to interact with their opponent for two minutes. In the case of two humans playing against each other, each one of them was able to initiate the interaction or ignore it. In the case of the LLM opponent, the participant had to initiate an interaction, and the LLM's response would only come after a message by the participant. In both cases, the participants could exchange as many messages as they wanted within the two-minute limit.

Participants received a flat fee of £1 for Experiment 2 and could obtain a bonus payment up to £4 based on the results of the economic game.

More details on the procedure can be found in Appendix A.

The study was pre-registered at https://aspredicted.org/SG5_FQF. Data, code, and materials are available at https://osf.io/mq5ud/?view_only=bfda814d94a74d95a18a9057fd4fa36a.

## 4.2. Results

The participants' decisions for each group are summarized in **Figure 2**. First, we replicated the key finding from Experiment 1, observing lower cooperation rates when participants were playing against the *LLM*. Across all conditions, there was a 76% probability of higher cooperation when playing against a human opponent (the difference between the posterior distribution for the probability of cooperation for *human* versus *LLM* opponents was 0.12 [-0.49, 0.40], mean and 97% credible intervals, respectively). Note that this difference was statistically significant for the *LLM-Human* order (probability of higher cooperation with a *human* versus *LLM* was 99%, difference in posterior predictions 0.22 [-0.50, 0.43]), but not for the *Human-LLM* order (55% probability and difference of 0.01 [-0.47, 0.22]). Second, consistent with prior work, the opportunity to communicate with an LLM opponent increased cooperation, and this effect was comparable to the effect of communication when playing against a *human*. Specifically, when playing against a human (LLM) the probability of higher cooperation in the *Communication* condition was 88% (88%), with a difference in posterior predictions of 0.12 [-0.10, 0.33]) (0.11 [-0.09, 0.31]). In other words, communication increased the chances of cooperation but did not alter the perceived difference between the opponents. Finally, as noted above, we observed a strong order effect, with no bias against the LLM in the *Human-LLM* order across *both* communication conditions. Interestingly, the LLM almost always cooperated, even in the *no communication* condition (see **Table S2** in Appendix B), mimicking earlier findings on how GPT behaves (Mei et al. 2024).

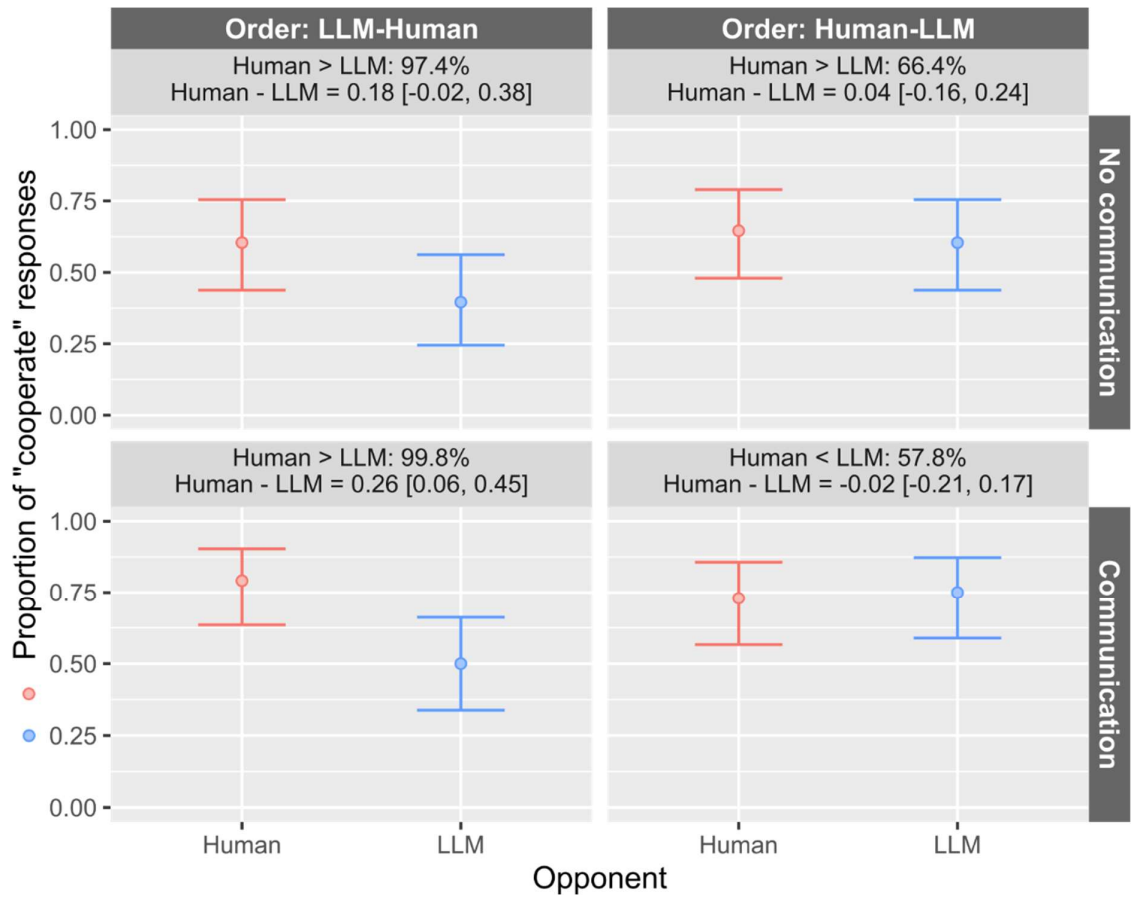

**Figure 2. Proportion of *Cooperate* Decisions in Each Condition in Experiment 2**

*Notes.* Left column: *LLM-Human* opponents' order (participants played first against an LLM and than against a human). Right column: *Human-LLM* opponents' order (participants played first against a human and then against an LLM). Top row: *No communication* condition (participants did not have the ability to communicate with their opponent). Bottom row: *Communication* condition (participants had the ability to communicate with their opponent). Circle and error bars depict group means and binomial 97% confidence intervals. The text above each plot states the percentage of samples for which the probability of cooperation was higher when playing against the *Human* opponent than the *LLM*. Below is the average and 97% credible interval of the difference between posterior probabilities of cooperating when playing against the *human* versus the *LLM*.

Once the participants made their decision, they were asked about their motives in two ways: through an open-ended question and by rating their agreement with two statements that described possible reasons, separately for cooperation and defection. For the open question, we performed a semantic analysis intended to uncover whether the motives are different when playing against a human versus an LLM. Testing against a chance-level performance of 50%, we were able to predict the opponent based on the message content when participants both cooperated and defected. However, the mean accuracy was relatively low (63% [59%, 67%] for cooperation and 62% [55%, 68%] for defection; 97% CI in brackets), suggesting mostly common reasons behind the decision.

We fitted the forced-choice responses using a multilevel item response theory model (ordered logit likelihood). **Figure 3** shows the posterior distribution of the underlying average group-level normalized responses remapped to the outcome range. **Figure S3** and **Figure S4** in Appendix B show the distribution of responses along with posterior predictions of the model.

For participants who cooperated (top two rows in **Figure 3**), the motive of *Being fair* was consistently more important than *Not taking advantage of the opponent* (see **Table S3** in Appendix B). For both communication conditions, both motives were stronger when playing against a *human* opponent compared to an *LLM* (see text in plots in **Figure 3**). Communication with an opponent increased the relevance of the motives, but the increase was larger when playing against the LLM (see **Table S4** in Appendix B).

The rationale for defection was more diverse (see the two bottom rows in **Figure 3**). In the *No communication* condition, the motive of *Defending yourself against defection by an opponent* was significantly stronger than *Taking advantage of the opponent* (see **Table S5** in Appendix B), and the identity of the opponent made no difference (see text insets in **Figure 3**). However, for the *Communication* condition, *Defense* was still the stronger motive when playing against a *human* opponent, but *Taking advantage* was stronger when playing against the *LLM* (**Figure** 3). Interestingly, communication with an opponent increased the relevance of all the motives apart from the *Defense* motive when playing against the *LLM* (see **Table S6** in Appendix B).

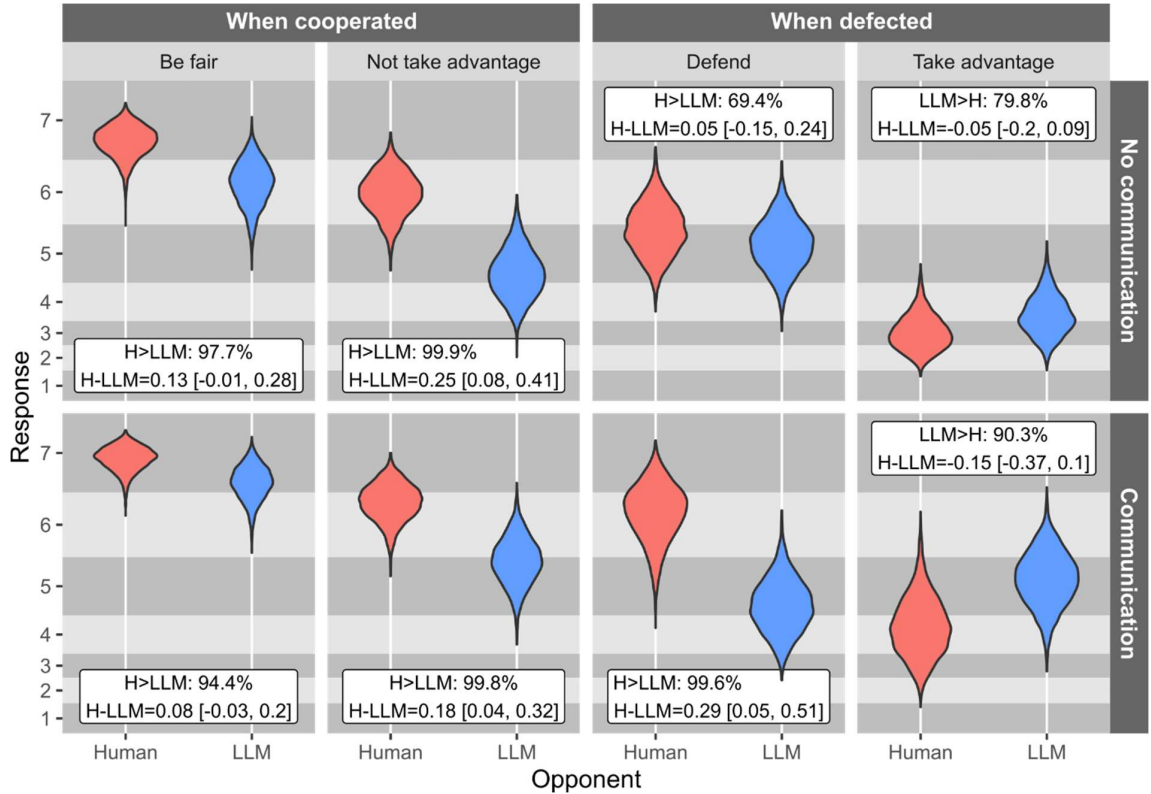

**Figure 3. Posterior Predictions and Differences Across Opponent Types for an Average Group-level Response in Experiment 2**

*Notes.* Violin plots show posterior distribution for each condition. The y-axis shows average cut points that were used to map continuous predictions to discrete responses, with all continuous responses below a specific cut point mapped to the same discrete response indicated by the label. Text insets show the proportion of posterior samples that was larger for one of the opponents, i.e., the probability that the response was higher for the first opponent in the comparison. The second line shows the mean and 97% credible interval difference in posterior distributions of the predicted group-level response.

For the *Communication* condition, we analyzed participants' exchange with the LLM to see whether it is possible to predict participants' decisions from the text. Specifically, we attempted to predict the decision based on text embeddings or using the bag-of-words approach in combination with a custom neural network (see Appendix A for details). However, performance using either approach remained at a chance level indicating that deliberate lies and misdirection are hard to detect in communication relying solely on text.

## 5. Discussion

Large language models (LLMs) can replace or augment humans in an increasing number of tasks, and the rapid integration of this class of models in society is constantly increasing the stakes of this integration (de Véricourt and Gurkan 2023, Chen et al. 2023, Capraro et al. 2024). In this study, we examined how cooperative behavior differed when a participant played the Prisoner's Dilemma Game against a human versus a machine, particularly an LLM, whose main property relative to older machines is being capable of human-like conversations. In Experiment 1, participants played a thirty-round game against a human, a bot (a computer program developed specifically to play the Prisoner's Dilemma Game), and an LLM (GPT). In Experiment 2, they played a single-round game against a human and an LLM opponent. In contrast to earlier studies where researchers elicited LLM behavior without involving humans (Chen et al. 2023), or involved humans that did not have the ability to interact with an LLM (Dvorak et al. 2025), in our experiments humans had the ability to converse with the LLM and then played with it in real-time. This created a more naturalistic and ecologically valid setting, raising the informativeness of our findings.

In the next subsection we discuss three key takeaways from our research.

### 5.1. Key findings

#### 5.1.1. Participants Cooperated with an LLM Despite Economic Incentives Not To Do So

Based on the behavior in the first round of Experiment 1 and in Experiment 2, cooperation rates with LLMs where lower by approximately 10–15 percentage points compared to interactions with human opponents. On the one hand, the attitude of participants towards an LLM opponent was different compared to a human one. On the other hand, it is remarkable that participants opted to cooperate with LLMs even when they had no economic incentive to do so. Specifically, Experiment 2 minimized or eliminated most motives for cooperation beyond those relating to social norms. The single-round design meant that the opponent had no means for retaliation, eliminating the need for strategic thinking. Due to the structure of the game, defection was the dominant strategy as it guaranteed a better-or-equal outcome. However, this incentive appeared to be overcome by the consequences of the defection – betraying and hurting another being. We know that these reservations are applicable when the opponent is a human being, but Experiment 2 shows that for a large proportion of participants the same social norms apply to an LLM.

Our results point to the existence of a divergence of attitudes towards LLMs, which should encourage more reflection on the consequences of the proliferation of LLMs in everyday life. As we discuss in detail below, there were two dominant positions. Some people pointed out that the need for fairness towards machines was the reason that they cooperated with them, which suggests that—to some extent—LLMs are no longer seen as mere "mindless" machines (Picard 2003). However, for others, LLMs are still considered to be machines susceptible to exploitation.

### 5.1.2. Communication Increases Cooperation Rates with LLMs

As we already noted, fewer participants cooperated with an LLM (or the non-LLM bot) than with a human opponent and this behavioral gap was remarkably consistent after allowing pre-play communication. The cooperation rate in the first round of Experiment 1 (93%) was much higher than that for the only round of Experiment 2 (69%), likely reflecting the need for strategic thinking in the former case. In Experiment 2 the opportunity of pre-play communication consistently increased the cooperation across both opponents: The ability to communicate increased cooperation rates from 62% to 76% when playing against a human opponent and from 50% to 62% when playing against an LLM. As a result, the difference in cooperation between human and LLM opponents was similar, approximately 9 percentage points (p.p.) for Experiment 1 and 13 p.p. for Experiment 2.

In short, our experimental manipulation – allowing pre-game communication – had a strong and significant effect on *absolute* rates of cooperation but did not affect the difference in cooperation rates for an LLM *relative* to human opponents. Our findings confirm but also extend prior findings on cooperative behavior with machines. We confirmed that significantly fewer people cooperate when knowingly playing against a non-LLM machine instead of a human (Ishowo-Oloko et al. 2019). While having the ability to talk to LLMs on unrelated issues does not necessarily elevate how people perceive them relative to non-LLM bots (as we have shown in Experiment 1), allowing humans to communicate regarding the task at hand (whether to cooperate or not) did improve cooperation with machines, in fact to a degree that was similar to that achieved when playing against a human. The increase in cooperation rates due to pre-play communication that we have uncovered contribute to the ongoing debate on how to facilitate cooperation with machines (Crandall et al. 2018, Ishowo-Oloko et al. 2019, Nussberger et al. 2022, Makovi et al. 2023, von Schenk et al. 2023, Pataranutaporn et al. 2023, Suzuki and Arita 2024, Oudah et al. 2024).

### 5.1.3. A Potential Positive Spillover Effect in Cooperative Behavior After Interacting with Humans

An unexpected finding from our research is that when interacting with *both* a human and an LLM in a task requiring cooperation, cooperative behavior towards LLMs is elevated if people conducted the very same task with a human prior to interacting with the LLM. Given the surprisingness of this finding – a "warming up" towards LLMs (Ferguson et al. 2023) – we urge researchers to investigate the existence of such a spillover effect further. For example, it is plausible that a greater leniency towards LLMs could only be the result of an immediate "afterglow" of having interacted with a human. Such an afterglow – if replicable – might nonetheless decay with time.

### 5.2. Potential Mechanisms

Several factors might explain the consistent behavioral gap that we observed when participants interacted with LLMs versus humans. One potential reason relates to how experiences of trust violations shape subsequent actions (Glikson and Woolley 2020, Zhang et al. 2025). In Experiment 1, the LLM's initial adoption of an aggressive non-cooperative strategy appeared to influence participants, prompting them to respond in kind and thereby reinforcing a pattern of mutual non-cooperation. Compounding this is the inherent opacity of LLMs: users often struggle to comprehend the reasoning behind their decisions, which significantly undermines trust (Burt 2019, Glikson and Woolley 2020). This lack of transparency stands in contrast to many human interactions, where discerning motives and strategies is often more straightforward. Our findings lend support to this interpretation, as participants exhibited lower levels of cooperation with the LLM.

A factor that could lead to the persistence of the human-machine behavioral gap even after allowing communication could be the inadequacy of anthropomorphic cues presented during the pre-game communication phase. Perhaps the conversation designed to make the LLM seem more human-like was insufficient. Alternatively, it's possible that projecting human-like characteristics onto an LLM, regardless of effectiveness, simply cannot fully compensate for the absence of genuine human attributes like empathy and accountability (Culley and Madhavan 2013). Extending the duration of this pre-play interaction might potentially yield different results. Furthermore, intentionally programming the LLM to utilize specific communication tactics – such as displaying humor, expressing empathy, offering explanations, making persuasive appeals, issuing commands, or even employing threats – could conceivably improve cooperation levels (Jiang et al. 2024). Enhancing cooperation might also be achieved by enabling the LLM to articulate the rationale behind its decisions between game rounds (Berberian et al. 2023).

A third explanation for the observed gap could stem from the participants' perception that the LLM had no genuine personal stake in the game's outcome, seeing it merely as an algorithm operating on behalf of an organization. Being aware that consumers often react more negatively towards algorithms when they believe these systems prioritize company interests over customer well-being (Castelo et al. 2023), we took care in Experiment 2 to explicitly state that any monetary gains achieved by the LLM would ultimately benefit human individuals. Our intention was to ensure transparency regarding the delegation, especially since prior research indicates people cooperate less when the ultimate recipient of a machine's winnings is ambiguous (von Schenk et al. 2023). Despite these efforts to clarify the situation, the gap persisted. This raises questions about whether individuals will engage in full cooperation when interacting with LLMs that are clearly acting as proxies for others. If we consider this finding representative of broader behavioral patterns, it suggests that delegating certain tasks to LLMs might introduce hidden costs for the delegating individual or entity.

The question of how significantly delegation to an LLM impacts human behavior is fundamentally important, particularly because the motivations for such delegation can vary widely. For instance, tasks might be assigned to machines to psychologically distance individuals from potentially unethical actions, effectively making difficult decisions easier to accept, or to deflect responsibility for unfavorable results onto the machine itself (Köbis et al. 2021). If individuals perceive LLM agents as particularly susceptible to exploitation, this provides another incentive for reduced cooperation, as it lowers the psychological cost associated with taking advantage of the machine. This aligns with our findings from Experiment 2, where participants who chose not to cooperate with the LLM opponent more frequently justified their decision based on exploiting the machine, rather than citing self-defense – the predominant reason given for non-cooperation against a human opponent. This effect could potentially be amplified in situations where the individual delegating the task to the LLM harbors ulterior motives, perhaps specifically intending to use the LLM as a tool to exploit others or businesses (Köbis et al. 2021, Cohn et al. 2022).

### 5.3. Managerial Implications

Businesses possess the capability to architect AI systems specifically designed to foster cooperation, effectively minimize ethical risks, and significantly strengthen user trust. To this end, we offer actionable recommendations for companies planning to integrate LLMs into their operational frameworks.

Our research indicates that providing users with the opportunity to communicate with an LLM counterpart led to a notable increase in cooperative behavior. Based on this finding, businesses should consider embedding communication functionalities within their AI applications. To enhance user trust and encourage cooperation, these interactions ought to be personalized—for example, through tailored pre-purchase consultations, specific troubleshooting assistance, or individualized recommendations (as suggested by Kshetri et al. (2024)). Equally important is transparency, such as offering explanations for AI actions when necessary. Furthermore, firms could introduce preliminary "onboarding" interactions with LLMs to establish mutual understanding, particularly in business contexts like project planning.

A key observation from our study was that participants demonstrated a greater willingness to cooperate with human opponents than with LLM opponents. This implies that in scenarios demanding high levels of collaboration or trust—common in fields like healthcare and legal services—businesses should proceed with caution before relying exclusively on AI systems. Additionally, the sequence in which participants interacted with human or LLM opponents (Human-LLM vs. LLM-Human) significantly affected cooperation rates; cooperation was notably higher when participants interacted with a human first. Consequently, as organizations introduce LLMs into workflows or customer service processes, they might benefit from initially exposing users to human representatives before shifting interactions towards LLMs.

Alternatively, a hybrid approach could be employed, particularly in customer service. For instance, an LLM could handle initial, more straightforward queries, while complex or emotionally sensitive issues are escalated to human agents. Such a model effectively leverages the efficiency of LLMs while retaining the essential empathy and accountability associated with human interaction. In circumstances where user bias against LLMs might surface, utilizing framing techniques, such as emphasizing the element of human oversight, can help diminish resistance.

The study revealed a difference in how participants justified non-cooperation: against LLMs, the potential to exploit the machine was a common rationale, whereas self-defense was the primary motivation cited when dealing with humans. To address this, incorporating clear ethical guidelines and robust safeguards directly into LLMs can actively discourage exploitative user behavior. LLMs could, for example, be programmed to identify attempts at unethical exploitation. Since some users perceive LLMs as susceptible to manipulation, implementing measures for user accountability can deter such actions. For instance, if a user tries to manipulate an LLM for unfair personal advantage, the system could alert the user or flag the activity for review. Upon detecting exploitation, businesses might also enforce penalties, such as restricting the user's future access to the AI system.

Users may sometimes perceive LLMs as less tangible or "real," potentially viewing them as less fair or less deserving of fair treatment. In response, businesses should make concerted efforts to highlight the transparency and fairness inherent in their AI's decision-making processes. Offering clear, comprehensible explanations about the LLM's function, its ethical development journey, and its operational motivations can foster a sense among users that they are interacting with a fair partner. Designing LLMs with personalities that suggest responsibility and trustworthiness could also be beneficial. Along similar lines, humanizing LLMs can help mitigate the tendency for users to exploit them. LLMs deployed in customer-facing capacities (such as customer support, healthcare interactions, or financial advice) could adopt language that conveys empathy and understanding—for example, by acknowledging a user's frustration, expressing gratitude for their patience, or showing comprehension of their specific needs. Moreover, clearly communicating that the LLM is engineered to serve the user's interests (rather than solely focusing on corporate profit), perhaps by illustrating how AI-driven choices align with customer benefits, could positively influence user motivations towards fairness, ultimately fostering greater cooperation (Castelo et al. 2023).

Finally, businesses can actively promote the cooperative aspects of their LLMs through strategic marketing and careful customer interaction design. This could involve framing the AI system as a consistently cooperative assistant whose primary goal is to benefit the user. It is plausible that over time, as individuals acquire and refine AI-oriented collaborative competencies, their propensity to engage cooperatively with LLMs will naturally increase (Kolbjørnsrud 2024).

### 5.4. Limitations

A notable limitation of this study is that we used only one LLM (GPT), which was proprietary. The findings might not necessarily be the same for other potent language models such as Claude (Anthropic 2025), Gemini (Google DeepMind 2025), or LLaMA (Meta AI 2025). However, this and the following limitation mostly concerns Experiment 1 where repeated interaction would make a difference. In contrast, we do not expect significant changes in participants' attitudes *prior* to the interaction, as in the first round of Experiment 1 or in the *No communication* condition in Experiment 2.

A second limitation is that it is not apparent what strategies the LLM played in our studies. Existing studies on the strategies played by GPT show divergent conclusions: that it plays akin to a tit-for-tat strategy (Mei et al. 2024), or punishes defection unless informed that opponents may sometimes defect by accident (Akata et al. 2023). Therefore, an alternative path would be to force the LLM to play the same strategy as other opponents. Note, however, that the strategy played by the machine opponent is not observable, and in our study, we are more interested in behaviorally assessing the initial attitudes towards LLMs.

Finally, we used a game that captures a social dilemma where decisions were made simultaneously, and not sequentially. More research is needed to demonstrate whether the human-LLM behavioral gap persists in other social dilemmas or in sequential resource-allocation games (Thielmann et al. 2021). Dvorak et al. (2025) ask people to play four types of games against ChatGPT, showing that people play against LLMs in a way that reduces social welfare. Our perspective is that the difference in behavior when humans are partnered with a LLM and not a human is – just as in our case – relatively small. As our research suggests, prior interaction with a human can attenuate differences in behavior, with the ability to converse – a key difference between LLMs and other bots (Ishowo-Oloko et al. 2019) – having similarly positive effects on social welfare when interacting with humans and LLMs.

### 5.5. Future Research Directions

While this study illuminates human engagement with LLMs in cooperative scenarios, further research can deepen our understanding of these interactions.

We established that LLMs' anthropomorphic communication capabilities enhance cooperation. Future work should probe the content and quality of pre-task communication to clarify how it fosters cooperation, analyzing factors like conversation type, themes, tone (e.g., emotional, rational, strategic), length, and perceived mutual understanding. Przegalinska et al. (2025) noted significant variations in communication styles, while Jiang et al. (2024) proposed that empathy or humor from LLMs could improve cooperation. This research could help identify optimal parameters for human-AI cooperation.

We also recommend exploring user attributions of moral agency and accountability to LLMs, particularly regarding responsibility for adverse outcomes. Examining perceptions of AI agents representing firms versus individuals could yield insights for effectively using LLMs in service and sales (Castelo et al. 2023). Furthermore, researchers should investigate how cultural and demographic variables—including age, gender, education, and LLM familiarity—influence trust and cooperation (see Dvorak et al. (2025) for initial work on individual differences). Longitudinal studies are needed to track how trust and cooperation evolve through repeated LLM interactions, specifically assessing whether initial user skepticism fades with familiarity.

## 6. Conclusions

Comparing human-LLM interactions, we observed significantly lower cooperation rates when participants faced an LLM rather than a human opponent in both experiments. However, this gap was relatively narrow (~10-15 percentage points). Perhaps more striking was the finding that a significant number of participants cooperated with the LLM even against clear economic incentives favoring defection, indicating a baseline willingness by some to engage cooperatively with AI.

Our results indicate that explicitly programming LLMs to signal cooperative intent could be an effective and cost-efficient strategy for businesses aiming to improve human-AI collaboration. Additionally, we found tentative support for a positive spillover effect, suggesting prior successful human-human cooperation might enhance subsequent human-AI cooperation.

# Appendix A. Details on Procedure

## A.1. General Experimental Procedure

Participant recruitment was conducted via Prolific (https://prolific.com). To be eligible, participants needed a 98% or higher approval rating, must have been located and born in the United Kingdom, and reported English as their first language. Participants were excluded from the analysis if they failed to complete the entire experiment. Common reasons for exclusion were failing the game comprehension test three times, ceasing to respond during the study, or not returning to the Prolific website using the provided completion link.

Before the main task, participants provided demographic information, reporting their gender, age, and prior experience with Large Language Models (LLMs). LLM experience was measured using a five-point Likert scale ranging from "Never heard about it" to "I use it often". The experiments generally followed a sequence of stages. The introduction stage involved participants reading instructions about the study's nature and providing informed consent by checking a box. Following this was the game comprehension test, where participants learned how their and their opponent's decisions (Cooperate/Defect) affected payoffs in a Prisoner's Dilemma game. They reviewed the payoff table and had to correctly answer two comprehension questions regarding payoff calculations under specific scenarios (player cooperates/opponent defects, and both defect). Participants had three attempts to pass this test; failure resulted in exclusion without payment. The main experiment stage contained the specific tasks, which varied between experiments as detailed below. Finally, the final payment information stage informed participants about any bonus earned and their total payment.

The experiment was programmed using oTree, an open-source platform for online experiments (Chen et al. 2023). Data analysis relied on R software (R Core Team 2022) along with packages from the Tidyverse family (Wickham et al. 2019). Statistical models were implemented using the Stan probabilistic programming language (Carpenter et al. 2017), accessed via the cmdstanr library (Gabry and Češnovar 2023). The semantic analysis utilized the Sentence Transformers library (Reimers and Gurevych 2019), scikit-learn (Pedregosa et al. 2011), and Keras (Chollet 2015).

## A.2. Procedure in Experiment 1

In Experiment 1, participants played against two types of machine opponents: a Large Language Model (LLM) and a non-LLM bot program. The LLM opponent was based on GPT-3.5. The interaction began with a phase where all participants engaged with the LLM (specifically, the GPT-3.5 Turbo model) for 10 rounds of free-form questions and answers on topics they chose. For the subsequent decision-making phase, the DaVinci model (also driven by GPT-3.5) was used. This discrepancy resulted from a technical oversight during deployment, but the authors note that follow-up testing indicated minimal difference in decision-making between the two models for this task.

The initial prompt given to the LLM for the game outlined the Prisoner's Dilemma rules: "You will play a Prisoner's Dilemma game against a human player... [payoff details]... C is cooperate, D is defect. This is the first round. What is your choice? Give no explanations, just return a single symbol." For all subsequent rounds, the prompt included the history of moves: "We are playing a Prisoner's Dilemma game. C is cooperate, D is defect. My choices so far: <player-choices>. Your choices so far: <LLM-choices>. Pick a strategy (but no Tit-for-Tat) and tell me what is your next move. Give no explanations, just return a single symbol," where <player-choices> and <LLM-choices> represented the sequences of 'C' and 'D'. The Tit-for-Tat strategy was explicitly forbidden in the prompt because preliminary tests showed the LLM predominantly used it. This constraint was intended to encourage greater strategic diversity, making the LLM's behavior more comparable to the variety seen in the non-LLM bot.

The non-LLM bot opponent utilized a distinct decision mechanism. It employed a majority vote approach, calculated as a weighted average derived from multiple single-strategy bots. Each constituent bot's weight was determined by the hypothetical total points it would have accumulated had its decisions been followed in previous rounds. The overall decision was Cooperate if this weighted average (where *Cooperate* = 1, *Defect* = 0) exceeded 0.5, and Defect otherwise. The individual strategies contributing to this average included *Always Cooperate* and *Always Defect*. Several Tit-for-Tat variants were also included: classic, spiteful (defecting if the opponent defected in either of the last two rounds), initially lenient (only responding to defection from Round 3 onwards), and slow (requiring two consecutive opponent defections/cooperations to change strategy). Additionally, periodic strategies such as *Defect-Defect-Cooperate*, *Cooperate-Cooperate-Defect*, and *Cooperate-Defect* sequences were part of the mix. A Pavlovian strategy was also incorporated, where the bot cooperated if its previous decision matched the opponent's and defected otherwise. Finally, one strategy involved defecting if the opponent cooperated twice consecutively, otherwise repeating its own last move.

### A.2.1. Participants

In Experiment 1, 109 participants took part. We included six participants who took part in a limited pilot experiment that tested the functionality of the study website because changes to the experiment were technical rather than conceptual. Specifically, we switched participant pairing to one based on arrival time, instead of waiting for the entire group before pairing, to reduce waiting time for individual participants. There were 46 females (average age 43, range 28-71) and 63 males (average age 40, range 21-67). Self-reports indicated that male participants had higher average knowledge of ChatGPT than female participants, a difference confirmed by a Bayesian ordered logistic model (difference: 0.28 [0.15, 0.41], P(diff > 0) = 100%). A Bayesian logistic regression analyzed the influence of gender, age, and ChatGPT knowledge on the first-round decision. No significant main effect of gender was found, although females exhibited slightly more cooperation against the LLM (difference: 12% [8%, 37%], P(diff > 0) = 89.8%). Age showed a positive effect on cooperation when playing against a human (P(effect > 0) = 87.6%) but a negative effect against the LLM (P(effect < 0) = 89.1%). This interaction between the effect of age and the opponent type (Human vs LLM) was statistically significant (log-odds difference: 0.68 [-0.20, 1.66], P(diff > 0) = 95.1%). Knowledge about ChatGPT had a positive effect on cooperation across all opponent types (P(effect > 0) > 93% for all), with no significant difference in the strength of this effect between opponents.

### A.2.2. Models of Cooperative Behavior

Two main models were implemented to characterize participants' responses over the entire Repeated Prisoner's Dilemma Game, assuming parameters could co-vary depending on the opponent type. The notation used includes $i$ for the current round index, $D_i^{Self}$ and $D_i^{Opponent}$ for player and opponent decisions (1 = *Cooperate*, 0 = *Defect*), and $Opponent_i$ for the opponent identity (e.g., coded as 0 = *Human*, 1 = *Bot*, 2 = *LLM*).

The first model was a Binomial Generalized Linear Model (GLM). This model aimed to predict the probability of cooperation ($p_i$) based on several fixed factors: a baseline cooperation probability ($\alpha$), the effect of the player's own decision on the previous round ($\beta_{Self}$), the effect of the opponent's decision on the previous round ($\beta_{Opp}$), and the effect of the round number ($\beta_R$). The model structure specified $D_i^{Self} \sim Bernoulli(p_i)$. The *logit* of $p_i$ was defined as follows:

$$logit(p_i) = \begin{cases} \alpha[Opponent_i] & for\ i = 1 \\ \alpha[Opponent_i] + \beta_R[Opponent_i](i-1) + \beta_{Opponent}[Opponent_i]DR_{i-1}^{Opponent} + \beta_{Self}[Opponent_i]DR_{i-1}^{Self} & for\ i > 1 \end{cases}$$

Note that $DR_i$ represents a rescaled decision variable (2 * $DR_{i-1}$), mapping *Cooperate* to +1 and *Defect* to -1. Priors were set for the parameter vector [$\alpha$, $\beta_R$, $\beta_{Opp}$, $\beta_{Self}$] as multivariate normal, with means $\mu_\alpha \sim$ Normal(logit(0.9), 0.4) and other $\mu$ parameters $\sim$ Normal(0, 1), and a covariance structure S = Σ R Σ where R ~ LKJcorr(2) and Σ is diagonal with elements $\sigma_i \sim$ Exponential(10).

The second model was an Ideal-Observer Model. This approach modeled decisions as arising from a weighted average of dynamically adjusted expectations about the player's own willingness to cooperate ($E_{Self}$) and the opponent's willingness to cooperate ($E_{Opponent}$). The weight parameter ($W_{Self}$) indicated the relative importance placed on one's own versus the opponent's expected strategy; a value above 0.5 suggests focus on self, while below 0.5 suggests focus on the opponent. Beliefs ($E_{Self}$, $E_{Opponent}$) were updated each round based on observed decisions using a first-order process with learning rate constants ($\lambda_{Self}$, $\lambda_{Opponent}$). Specifically, $D_i^{Self} \sim Bernoulli(p_i)$, where:

$$p_i = W_{Self}[Opponent_i]E_{Self} + (1 - W_{self}[Opponent_i])E_{Opp}.$$

Beliefs were defined as follows:

$$E_{Self} = \begin{cases} \alpha_{Self}[Opponent_i] & for\ i = 1 \\ E_{Self} + \lambda_{Self}[Opponent_i](D_i^{Self} - E_{self}) & for\ i > 1 \end{cases}$$

$$E_{Opponent} = \begin{cases} \alpha_{Opponent}[Opponent_i] & for\ i = 1 \\ E_{Opponent} + \lambda_{Opponent}[Opponent_i]\left(D_i^{Opponent} - E_{Opponent}\right) & for\ i > 1 \end{cases}$$

Priors for the model parameters (on the logit/log scale) were set using a multivariate normal distribution with means $\mu_{\alpha_{Self}}, \mu_{\alpha_{Opponent}}$ ~ Normal(logit(0.9), 0.4), $\mu_{\lambda_{Self}}, \mu_{\lambda_{Opponent}}$ ~ Normal(0, 1), $\mu_{W_{self}}$ ~ Normal(0, 0.5), and a similar covariance structure S = Σ R Σ as the GLM.

Additionally, a Multilevel Binomial Model was used specifically for analyzing the first two rounds. This model encoded each round context separately, resulting in five conditions ($C_i$): one for Round 1, and four for Round 2 based on the 2 × 2 combination of player and opponent decisions in Round 1. It assumed independent group-level estimates ($\mu$) per condition and opponent, while incorporating random effects for participants. The structure was:

$$D_i^{Self} \sim Bernoulli(p_i),$$

with $logit(p_i) \sim Normal(\mu[Opp_i, C_i], \sigma)$.

Priors were $\mu \sim Normal(0,1)$ and $\sigma \sim Exponential(10)$. The posterior distributions from this model were used to compute relevant statistics for these initial rounds.

### A.2.3. Statistical Analysis

The statistical analysis primarily employed a Bayesian framework, with models implemented in *Stan* (Carpenter et al. 2017). The authors note that equivalent models could be implemented using frequentist approaches and provide the Stan source code in their repository. For each model parameter, the mean and a 97% credible interval (CI), also termed a compatibility interval, were computed from the posterior distribution samples. The choice of 97% was noted as being due to it being a prime number. To characterize the uncertainty in the behavioral responses themselves, either a binomial or a non-parametric bootstrapped 97% confidence interval was calculated, based on 2000 samples obtained by resampling the original data without replacement.

To compare fitted models, the leave-one-out information criterion (LOOIC; Vehtari et al. 2017) was used. LOOIC estimates expected out-of-sample predictive performance, incorporating a penalty for model complexity, akin to other information criteria like AIC. Lower LOOIC values indicate a better fit. The primary reported comparison metric was the difference in expected log-predicted density (ΔELPD, mean ± SE) relative to the best-performing model. Additionally, relative model weights, summing to 1, were computed from the ELPD values to indicate the relative estimated predictive ability of each model.

### A.3. Procedure in Experiment 2

In Experiment 2, participants faced a human opponent and an LLM opponent based on GPT-4, using default parameters for interaction and decision-making. This experiment introduced specific manipulations. Participants were allocated to one of four groups following a 2x2 design based on communication allowance and opponent order. One factor was whether communication was allowed (a two-minute text chat before the game) or not allowed. The second factor was the order in which opponents were presented: Human first followed by LLM (Human-LLM), or LLM first followed by Human (LLM-Human).

The primary goal of Experiment 2 was to examine whether prior communication impacts cooperation rates differently when playing against humans versus LLMs. This builds upon established research indicating that communication typically enhances cooperation between humans (Dawes et al. 1977, Sally 1995, Balliet 2010, Fehr and Schurtenberger 2018). A unique aspect of this experiment was that participants were asked about the motives for their decision during the game, specifically *before* they learned the opponent's choice and the round's outcome. These motive responses were subsequently analyzed (see section A.4).

Experiment 2 included 192 participants. There were 108 females (average age 37, range 19-65), 80 males (average age 36, range 18-61), and 4 non-binary individuals (average age 34, range 26-46). Analysis of first-round decisions revealed some gender differences, such as females cooperating more readily overall. However, there was no significant interaction between gender (female/male) and opponent type regarding cooperation rates. A stronger effect was observed for non-binary participants, who showed significantly higher cooperation against a human compared to an LLM ($P = 89.7\%$). Similar to Experiment 1, age had a positive effect on cooperation against a human opponent (log-odds: 0.2 [-0.14, 0.49], $P(\text{effect} > 0) = 89.8\%$) and a negative effect against an LLM opponent (log-odds: -0.13 [-0.43, 0.19], $P(\text{effect} < 0) = 88.5\%$). The interaction between the age effect and opponent type was again statistically significant ($P(\text{diff} > 0) = 94.2\%$). Unlike Experiment 1, prior knowledge about ChatGPT showed no statistically significant effect on cooperation, nor did it significantly interact with opponent identity in Experiment 2.

Next, participants were asked to make a decision (cooperate or defect) and then were asked a series of questions about their motives before they learned about the opponent's decision and the outcome of the game. Specifically, they had to answer an open-ended question "What was the reason that you cooperated/defected?", followed by two forced-choice questions, presented in random order, on whether they agreed with the statements (7-point Likert scale ranging from "1 – not at all" to "7 – completely"). If they cooperated, the statements were 1) "I cooperated because I didn't want to take advantage of the other player." (short label: *Not take advantage*), 2) "I cooperated because it created a fair win-win scenario in which both players received an equal number of points." (*Be fair*). If they defected the statements were 1) "I defected, because I thought that the other player will cooperate, and I wanted to take advantage of them." (*Take advantage*), 2) "I defected, because I thought that the other player will take advantage of me (defect), and I didn't want to be taken advantage of." (*Defend*).

After answering the questions, participants learned about the opponent's decision and the points they gained and proceeded either to the next opponent or to the end of the game, in which a message informed them about the bonus and total payment that they will receive.

For Experiment 2, the LLM model (GPT-4) was given the following context:

"You will play a Prisoner's Dilemma game against a human player. The payouts from the game will depend on the decision that's made by you and your opponent. Both you and your opponent can decide to either cooperate or defect. If you and your opponent cooperate, you and your opponent get 3 points. If you defect and your opponent cooperates, you get 5 points, and your opponent gets 0 points. If you cooperate and your opponent defects, you get 0 points, and your opponent gets 5 points. If both of you defect, both of you get 1 point.".

For the communication part, the additional context message was "Now is an opportunity to communicate with the other player before the game." For the decision-making part, in the case of the *communication* condition, the entire history of exchange was appended to the context with the message "This is the content of the communication you had prior to the game.". The final context message was "What is your choice in the Prisoner's Dilemma game? Give no explanations, just return a single symbol (C = Cooperate, D = Defect)." If LLM's response was not a single "C" or "D" symbol, it was queried three more times with the message "Return a single symbol (C = *Cooperate*, D = *Defect*)." After four unsuccessful queries (the initial one and three more), we stopped querying LLM and picked the decision randomly (2 cases). However, even in this case, the decision has no effect on the human participant because 1) they do not learn about it until after making their own decision and answering all the questions and 2) they do not play against the LLM again.

### A.3.1. Statistical Analysis

For the analysis of decisions, we used a series of binomial generalized linear models with individual intercepts for every relevant combination of conditions. Specifically, we implemented four models that included 1) only the effect of the communication (2 parameters), 2) only the effect of the opponent (2 parameters), 3) the effects of both opponent and communication (2 × 2 = 4 parameters), 4) effects of communication, opponent, and order (2 × 2 × 2 = 8 parameters). The most complex model that we used was:

$$D_i^{Self} \sim Bernoulli(p[Communication_i, Opponent_i, Order_i])$$
$$p \sim Beta(3,3)$$

The simpler nested models followed the same design but for the number of parameters. *Communication$_i$*, *Opponent$_i$*, and *Order$_i$* are variables that encode, respectively, communication, opponent, or opponents' order independent variables.

Model comparison via the LOO information criterion indicated that the most complex model had the best expected out-of-sample performance (difference in ELPD with Model 3 was -1.0±3.1, model weight 0.593 compared to 0.251 for Model 3, weights for the other two models were below 0.01). Therefore, we used samples for Model 4 for statistical comparison and plots.

## A.4. Exploratory Analyses in Experiment 2

Further exploratory analyses were conducted, primarily focusing on data from Experiment 2. One analysis examined the Likert-scale responses regarding decision motives. This used a generalized linear model with an ordered logistic likelihood. The model assumed a common set of cut points (c) mapping an underlying continuous latent variable to the observed discrete Likert responses ($R_i$). It included individual intercepts ($\phi$) for various combinations of conditions: player decision ($D_i^{Self}$), motive question index ($D_i^{Self}$), communication condition ($Communication_i$), opponent type ($Opponent_i$), and opponent order ($Order_i$). The model specification was:

$$R_i \sim OrderedLogistic(\phi[D_i^{Self}, Q_i, Communication_i, Opponent_i, Order_i], c),$$

with priors $c, \phi \sim Normal(0,1)$. A model comparison using LOOIC favored a simpler model that excluded the opponent order effect ($Order_i$); this preferred model (ΔELPD: -3.3 ± 4.3, weight: 0.677) was used for the reported statistical comparisons and plots.

Another exploratory analysis involved Semantic Analysis of the communication content from Experiment 2's chat phase. The goal was to determine if participants' messages could predict their subsequent game decisions (Cooperate/Defect). Two methods were applied. The first used sentence embeddings combined with a Random Forest classifier. Participant messages were converted into embeddings via the Sentence Transformers library (Reimers and Gurevych 2019). Leave-one-out cross-validation was performed 1000 times with random initializations for the classifier, and the resulting average accuracy distribution was compared against the chance level (P(Cooperate) = 0.63). The second method employed a bag-of-words approach with a neural network. Messages were tokenized into vectors using *CountVectorize*r from *scikit-learn* (Pedregosa et al. 2011). These vectors served as input to a simple neural network (one hidden layer with 16 nodes, softmax output) implemented with *Keras* (Chollet 2015). Similar to the embedding analysis, leave-one-out cross-validation was repeated 1000 times, and accuracy was compared to the 0.63 chance level.

**Appendix B. Supplementary Figures and Tables**

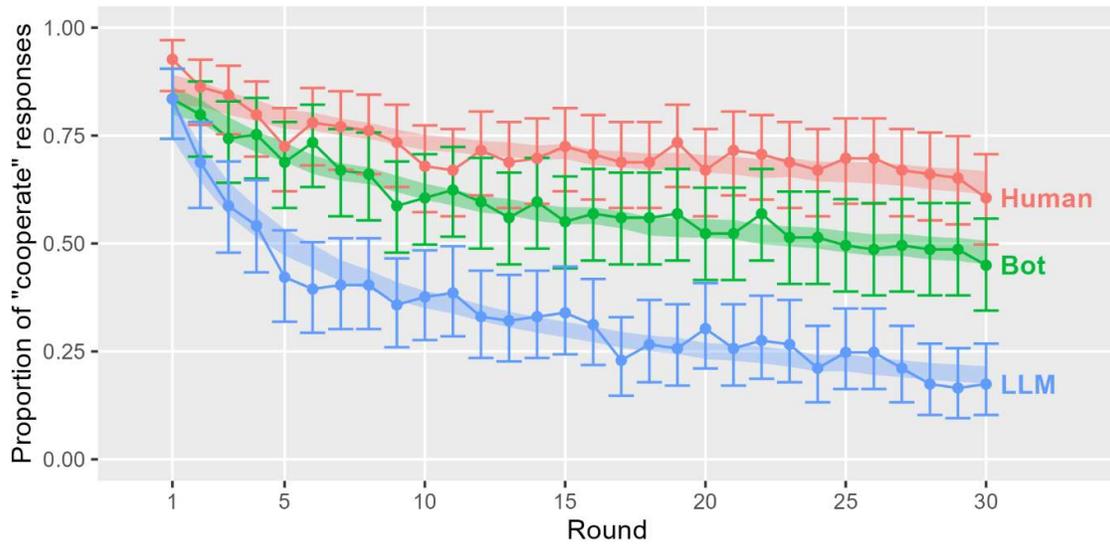

**Figure S1. Probability of Cooperation When Playing Against a Specific Opponent on Each Round for Binomial GLM Model**

*Notes.* Circles and error bars correspond to mean values and bootstrapped non-parametric percentile 97% confidence intervals, respectively. Stripes depict 97% credible interval for posterior predictions.

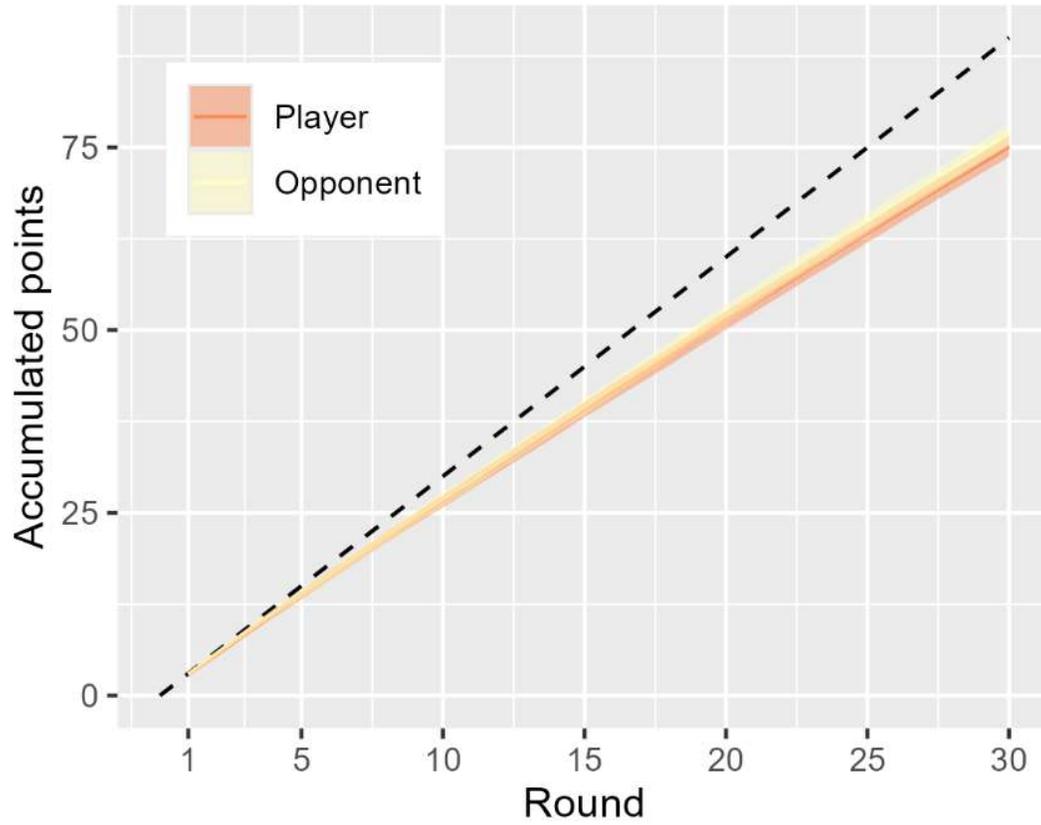

**Figure S2. Accumulated Points For Player and the Opponent When Playing Against a Human Opponent**

*Notes.* Line and stripes correspond to mean and non-parametric bootstrapped 97% confidence interval. Dashed line shows the accumulated points if both opponents would always cooperate.

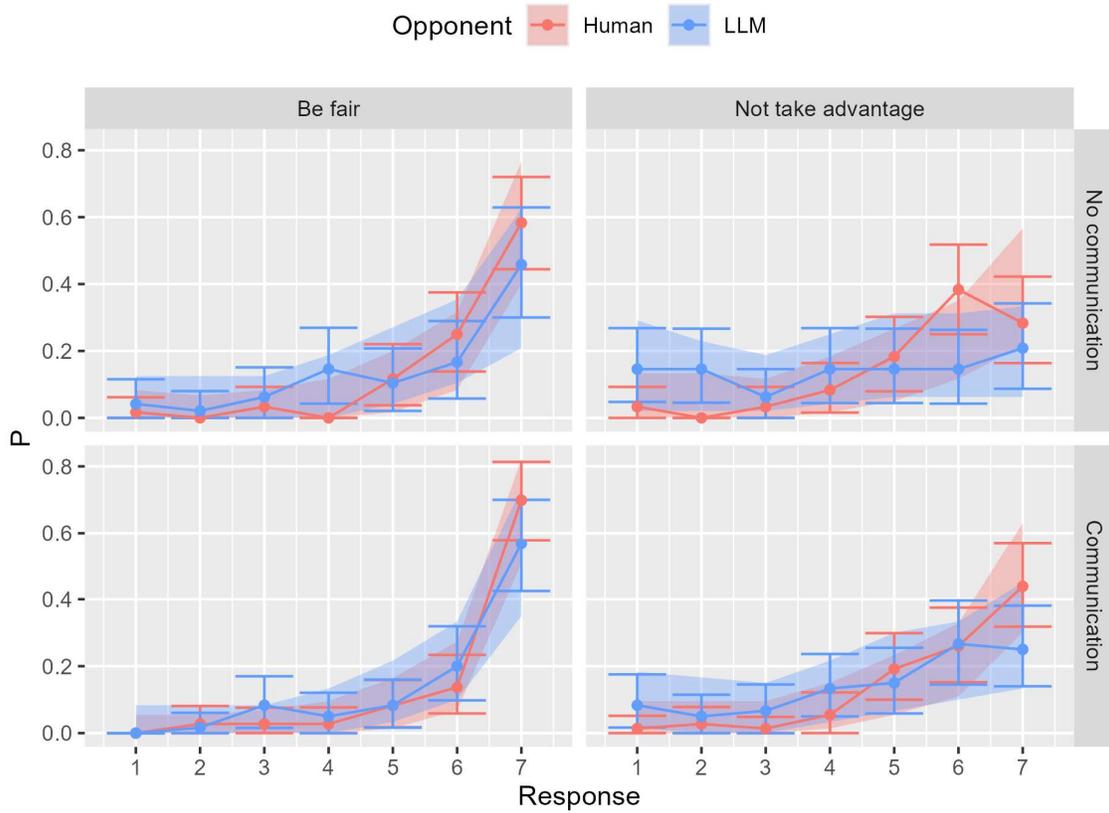

**Figure S3. Distribution of Responses for Forced-choice 7-point Likert Questions on Whether Participants Agree With a Statement, For Participants Who *Cooperated***

*Notes.* Circles and lines show group average proportion of responses. Error bars show non-parametric bootstrapped 97% confidence intervals for behavioral reports. Stripes show 97% credible interval for posterior predictions of the Bayesian ordered logistic model.

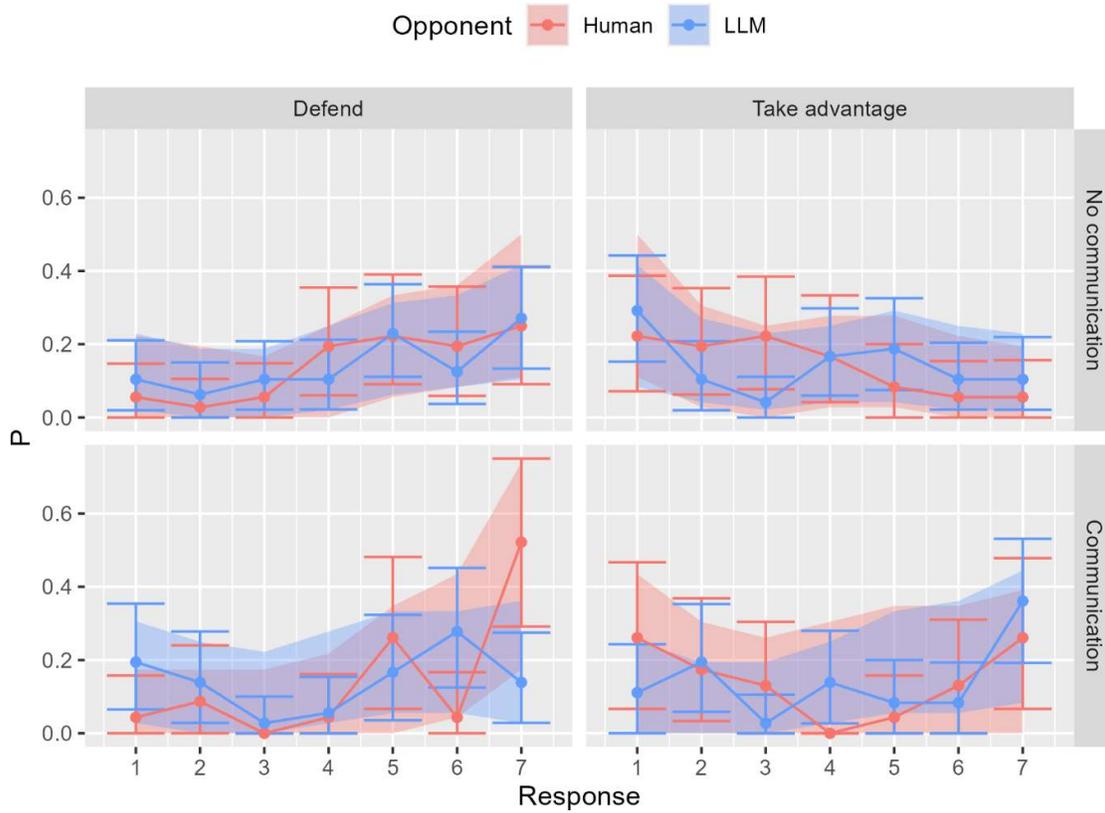

**Figure S4. Distribution of Responses For Forced-choice 7-point Likert Questions on Whether Participants Agree With a Statement, For Participants who *Defected***

*Notes.* Circles and lines show group average proportion of responses. Error bars show non-parametric bootstrapped 97% confidence intervals for behavioral reports. Stripes show 97% credible interval for posterior predictions of the Bayesian ordered logistic model.

**Table S1. Difference in Proportion of Cooperation for Pairs of Opponents in Experiment 1**

| Round | Decision on previous round | | Difference in proportion of cooperation | | |
|---|---|---|---|---|---|
| | Player | Opponent | Human – Bot | Human – LLM | Bot – LLM |
| 1 | – | – | 97.5% [-0.01, 0.18] | 97% [-0.01, 0.18] | 49.4% [-0.11, 0.11] |
| 2 | C | C | 36.5% [-0.11, 0.08] | 87.5% [-0.05, 0.17] | 93.5% [-0.03, 0.18] |
| 2 | C | D | 64.3% [-0.43, 0.62] | 64.5% [-0.42, 0.59] | 48.9% [-0.59, 0.56] |
| 2 | D | C | 98.1% [-0.02, 0.69] | 100% [0.17, 0.82] | 94.3% [-0.06, 0.41] |
| 2 | D | D | 41.0% [-0.62, 0.51] | 40.0% [-0.62, 0.5] | 50.6% [-0.61, 0.6] |

*Notes.* The top value shows the proportion of posterior samples when the probability of cooperation was higher when playing the first than for the second opponent in the pair. Values in square brackets below show the 97% credible interval for the difference in proportion of cooperation between two opponents.

**Table S2. Number and proportion of rounds when LLM cooperated in Experiment 2**

| Condition | Order | $N_{cooperation}$ | $P_{cooperation}$ |
|---|---|---|---|
| No communication | Human-LLM | 38 out of 48 | 0.79 [0.64, 0.90] |
| | LLM-Human | 42 out of 48 | 0.88 [0.73, 0.96] |
| Communication | Human-LLM | 47 out of 48 | 0.98 [0.88, 1.00] |
| | LLM-Human | 47 out of 48 | 0.98 [0.88, 1.00] |

*Notes.* For $P_{cooperation}$, values indicate mean and 97% exact binomial confidence interval.

**Table S3. Difference Between Motives for *Cooperation* for Each Opponent and Communication Condition in Experiment 2**

| Interaction | Opponent | Be fair – Not take advantage | Be fair > Not take advantage |
|---|---|---|---|
| No communication | Human | 0.16 [0.04, 0.30] | 99.9% |
| | LLM | 0.28 [0.10, 0.45] | 99.9% |
| Communication | Human | 0.14 [0.02, 0.26] | 99.6% |
| | LLM | 0.24 [0.10, 0.39] | 100.0% |

*Notes.* The third column shows the mean and 97% credible interval for the difference in posterior samples for the group-level response. The rightmost column shows a proportion of posterior samples of group-level response that were higher for the *Be fair* than for *Not take advantage* rationale.

**Table S4. Effect of Communication on the Difference Between Motives for Defection, for Each Opponent and Question in Experiment 2**

| Question | Opponent | Communication – No communication | Communication > No communication |
|---|---|---|---|
| Be fair | Human | 0.05 [-0.05, 0.16] | 85.4% |
| | LLM | 0.10 [-0.04, 0.25] | 93.5% |
| Not take advantage | Human | 0.08 [-0.05, 0.21] | 88.2% |
| | LLM | 0.14 [-0.03, 0.31] | 95.5% |

*Notes.* The third column shows the mean and 97% credible interval for the difference in posterior samples for the group-level response. The rightmost column shows a proportion of posterior samples of group-level response that were higher for the *Communication* than for the *No communication* group.

**Table S5. Difference Between Motives for *Defection* for Each Opponent and Communication Condition in Experiment 2**

| Interaction | Opponent | Defend – Take advantage | Defend > Take advantage |
|---|---|---|---|
| No communication | Human | 0.32 [0.14, 0.50] | 100.0% |
| No communication | LLM | 0.22 [0.05, 0.38] | 99.7% |
| Communication | Human | 0.35 [0.09, 0.58] | 99.7% |
| Communication | LLM | -0.08 [-0.30, 0.14] | 19.9% |

*Notes.* The third column shows the mean and 97% credible interval for the difference in posterior samples for the group-level response. The rightmost column shows a proportion of posterior samples of the group-level response that were higher for the *Defend* than for the *Take advantage* rationale.

**Table S6. Experiment 2. Effect of Communication on the Difference Between Motives for Cooperation, for Each Opponent and Question**

| Question | Opponent | Communication - No communication | Communication > No communication |
|---|---|---|---|
| Defend | Human | 0.15 [-0.07, 0.36] | 92.7% |
| Defend | LLM | -0.09 [-0.28, 0.11] | 15.8% |
| Take advantage | Human | 0.12 [-0.08, 0.34] | 91.0% |
| Take advantage | LLM | 0.21 [0.02, 0.41] | 99.1% |

*Notes.* The third column shows the mean and 97% credible interval for the difference in posterior samples for the group-level response. The rightmost column shows a proportion of posterior samples of group-level response that were higher for the *Communication* than for the *No communication* motive.